\documentclass[11pt]{article} 
\usepackage{epsfig,wrapfig} 
\usepackage{cite}  
\usepackage{graphics}  
\def\eps{\epsilon}

\def\as{\alpha_s}

\def\nn{\nonumber}

\begin{document} 
\unitlength1cm 
\begin{titlepage} 
\vspace*{-1cm} 
\begin{flushright} 
ZU--TH 02/06\\
January 2006
\end{flushright} 
\vskip 3.5cm 

\begin{center} 
{\Large\bf Towards $e^+e^-\to 3$\,jets at NNLO by sector decomposition}
\vskip 1.cm 
 {\large G.~Heinrich}
\vskip .7cm 
{\it  Institut f\"ur Theoretische Physik, Universit\"at Z\"urich,
Winterthurerstrasse 190,\\ CH-8057 Z\"urich, Switzerland} 
\vskip .4cm 
\end{center} 
\vskip 2.6cm 

%VERSION \today

\begin{abstract} 
A method based on sector decomposition has been developed 
to calculate the double real radiation part of the process 
$e^+e^-\to 3$\,jets at next-to-next-to-leading order. 
It is shown in an example that the numerical cancellation 
of soft and collinear poles works well. The method is flexible to
include an arbitrary measurement function in the final Monte 
Carlo program, such that it allows to obtain differential
distributions for different kinds of observables. 
This is demonstrated by showing 3--, 4-- and 5--jet rates 
at order $\alpha_s^3$ for a subpart of the process.

\end{abstract} 
\vfill 
\end{titlepage} 
\newpage 

\renewcommand{\theequation}{\mbox{\arabic{section}.\arabic{equation}}} 

\section{Introduction}
\setcounter{equation}{0}
Experiments at LEP have shown that the measurement of jet rates and 
shape observables in $e^+e^-$ collisions allow for very stringent
tests of the Standard Model, in particular of predictions 
relying largely on Quantum Chromodynamics
(QCD)\cite{Dissertori:2003pj}, allowing for example a very precise 
determination of the strong coupling constant $\alpha_s$. 
A precise knowledge of $\alpha_s$ in turn is of major importance 
at hadron colliders, especially at the LHC.
However, the measurements from jets and shapes in $e^+e^-$ collisions,
although being very precise, have not been included in the 
world average value for $\alpha_s$, because it is based only on 
measurements where next-to-next-to leading order (NNLO) 
theory predictions are available\cite{Bethke:2004uy}, while   
for $e^+e^-\to 3$\,jets, full NNLO predictions do not exist yet.
 
A future International Linear Collider will allow for precision
measurements at the per-mille level, which offer the possibility 
of a determination of  $\alpha_s$ 
with unprecedented precision. 
However, this will only be possible if the theoretical error 
can keep up with such a precision. As the present error on 
the NLO prediction for $e^+e^-\to 3$\,jets is dominated by scale 
uncertainties\cite{Kramer:1986mc,Ellis:1980wv,Fabricius:1981sx,Kunszt:1989km}, 
the calculation of the NNLO corrections to this process will surely
lead to an important gain in precision. 

After the virtual two-loop corrections entering this calculation
have become available\cite{Garland:2001tf,Garland:2002ak,Moch:2002hm}, 
the bottleneck now is given by the real radiation part 
where up to two partons can become theoretically unresolved (soft
and/or collinear), leading to infrared singularities upon phase 
space integration. 
These singularities have to be subtracted and cancelled with the ones 
from the virtual contributions before a Monte Carlo program 
can be constructed. At NNLO, the infrared singularities can be 
entangled in a complicated way, which renders the extraction of the
poles a formidable task.
Two different approaches can be followed to achieve this task:
\begin{enumerate}
\item  
Construction  of a subtraction scheme where 
%the subtraction terms approach the singular behaviour of the 
%matrix element in all unresolved limits; these 
the subtraction terms are integrated 
analytically in $D=4-2\eps$ dimensions over the 
unresolved phase space, thus extracting the poles in $1/\eps$. 
The main advantages of this approach are the following: 
It allows maximal (i.e. analytical) control over the pole terms,  
and it insures a minimal number of subtraction terms, as the latter 
are constructed manually by considering all physical situations
where a singular configuration is approached. 
The drawbacks of this method are given by the fact that 
constructing such a scheme is a highly non-trivial and tedious task, 
especially in view of the fact that it is different for each 
colour structure. Further, the analytic integration over subtraction
terms may become impossible  when applying the method to other
processes where several mass scales are involved.
\item  Sector decomposition, where the poles are isolated 
by an automated routine and the pole coefficients are integrated 
numerically. The advantages of this approach reside in the fact that 
the extraction of the infrared poles is algorithmic, 
being the same for all colour factors,
%thus minimizing the manual work and guaranteeing applicability to
%other processes,
and that the subtraction terms can be arbitrarily 
complicated as they are integrated only numerically.  On the other
hand, the algorithm which isolates the poles 
increases the number of original functions
and in general does not lead to the 
minimal number of subtraction terms, 
thus producing rather large expressions. 
\end{enumerate}
Approach 1. has been pursued by several groups 
in different variations\cite{Kosower:2002su,Kosower:2003cz,Weinzierl:2003fx,
Weinzierl:2003ra,Kosower:2003bh,Gehrmann-DeRidder:2003bm,Gehrmann-DeRidder:2004tv,Kilgore:2004ty,Frixione:2004is,
Gehrmann-DeRidder:2004xe,Gehrmann-DeRidder:2005hi,Gehrmann-DeRidder:2005aw,Somogyi:2005xz,Gehrmann-DeRidder:2005cm},
and the implementation of the method based on antenna 
subtraction\cite{Gehrmann-DeRidder:2004xe,Gehrmann-DeRidder:2005cm}  into a Monte Carlo program 
is presently underway\cite{GGGH06}.  
The sector decomposition approach has seen a very rapid development recently. 
Sector decomposition is a general method to disentangle overlapping singularities 
in parameter space, originally used by K.~Hepp\cite{Hepp:1966eg} for overlapping
ultraviolet singularities. It has been very successfully applied to 
various types of multi-loop integrals 
since\cite{Roth:1996pd,Binoth:2000ps,Binoth:2003ak,Denner:2004iz,Heinrich:2004iq,Czakon:2004wm,Anastasiou:2005pn}. 
Its application to NNLO phase space integrals, 
first proposed in\cite{Heinrich:2002rc}, 
already lead to a number of promising results\,
\cite{Gehrmann-DeRidder:2003bm,Anastasiou:2003gr,Binoth:2004jv,Anastasiou:2004qd,Anastasiou:2005qj,Anastasiou:2005pn}.

The present paper deals with the application of sector decomposition to the 
double real radiation part of the process $e^+e^-\to 3$\,jets at NNLO, 
which involves  subprocesses of the type $\gamma^*\to 5$\,partons. 
The matrix elements for the processes $\gamma^*\to q\bar{q}ggg$ and
$\gamma^*\to q\bar{q}q^\prime\bar{q}^\prime g$ are huge, such that 
the calculation also involves non-trivial book-keeping and file handling tasks, 
which are not addressed in this article. 
The intention of this paper is to show that a method has been developed 
which can deal with $1\to 5$ parton processes efficiently, such that 
the construction of a fully differential Monte Carlo program for 
the process $e^+e^-\to 3$\,jets at NNLO is merely a matter of 
putting pieces together, which will be left to a future publication. 
Therefore, we only consider one sample topology (including the full tensor
structure) as part of the full 
matrix element. 
For this topology, we first calculate the fully inclusive integral 
over the 5-parton phase space, leading to poles up to $1/\eps^4$. 
In order to prove the correctness of the result, 
we also calculate all possible cuts of this diagram 
with less than five particles in the final state. 
For the contribution from $\gamma^*\to 4$ partons, we use the result 
obtained in \cite{Binoth:2004jv} by sector decomposition. 
The KLN theorem\cite{Kinoshita:1962ur,Lee:1964is} 
guarantees that the sum of all possible cuts of the UV renormalised
diagram is finite. This is demonstrated in section 2. 
However, the method is not limited to calculate only inclusive cross 
sections. As the singularities are disentangled by an algebraic 
algorithm, the inclusion of an arbitrary (infrared safe) measurement 
function -- at the stage of the numerical evaluation of the finite 
functions produced by sector decomposition -- does not present a 
problem. This is shown in section 3. As an illustration of the action 
of the measurement function, the ${\cal O}(\alpha_s^3)$ 
3--, 4-- and 5--jet
rates are shown for the sample matrix element as a function of the 
cut parameter $y^{\rm cut}$ within the JADE algorithm\cite{Bethke:1988zc}. 
Section 4 contains the conclusions. Details of the calculation are given in 
appendix A.

\section{Cancellation of Divergences}
\setcounter{equation}{0}
\label{sec:div}
As explained above, the method presented here addresses the 
main difficulty in calculating the real radiation part 
of $e^+e^-\to 3$\,jets at NNLO, which is  the isolation and 
subtraction of the infrared poles which occur when integrating the 
squared amplitude over the phase space for $\gamma^*\to 5$ partons. 

\begin{figure}[htb]
%\begin{picture}(80,60)
\begin{center}
\epsfig{file=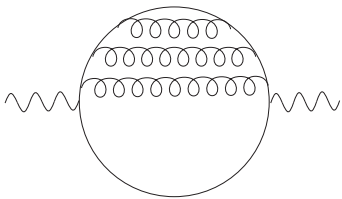,height=3.cm}
\end{center}
\vspace*{-0.5cm}
\caption{The ladder diagram\label{fig1}}
\end{figure}

In order to check the correctness of the results for the integrals 
over the $1\to 5$ particle phase space, one can exploit 
the fact that the sum over all cuts of a given 
(UV renormalised) topology must be infrared finite. 
In order to demonstrate these cancellations, let us consider 
as an example the diagram 
depicted in Fig.~\ref{fig1}, occurring in the part $\sim C_F^3$ 
of the squared amplitude for $e^+e^-\to 3$\,jets at NNLO.

\begin{figure}[htb]
%\begin{center}
\epsfig{file=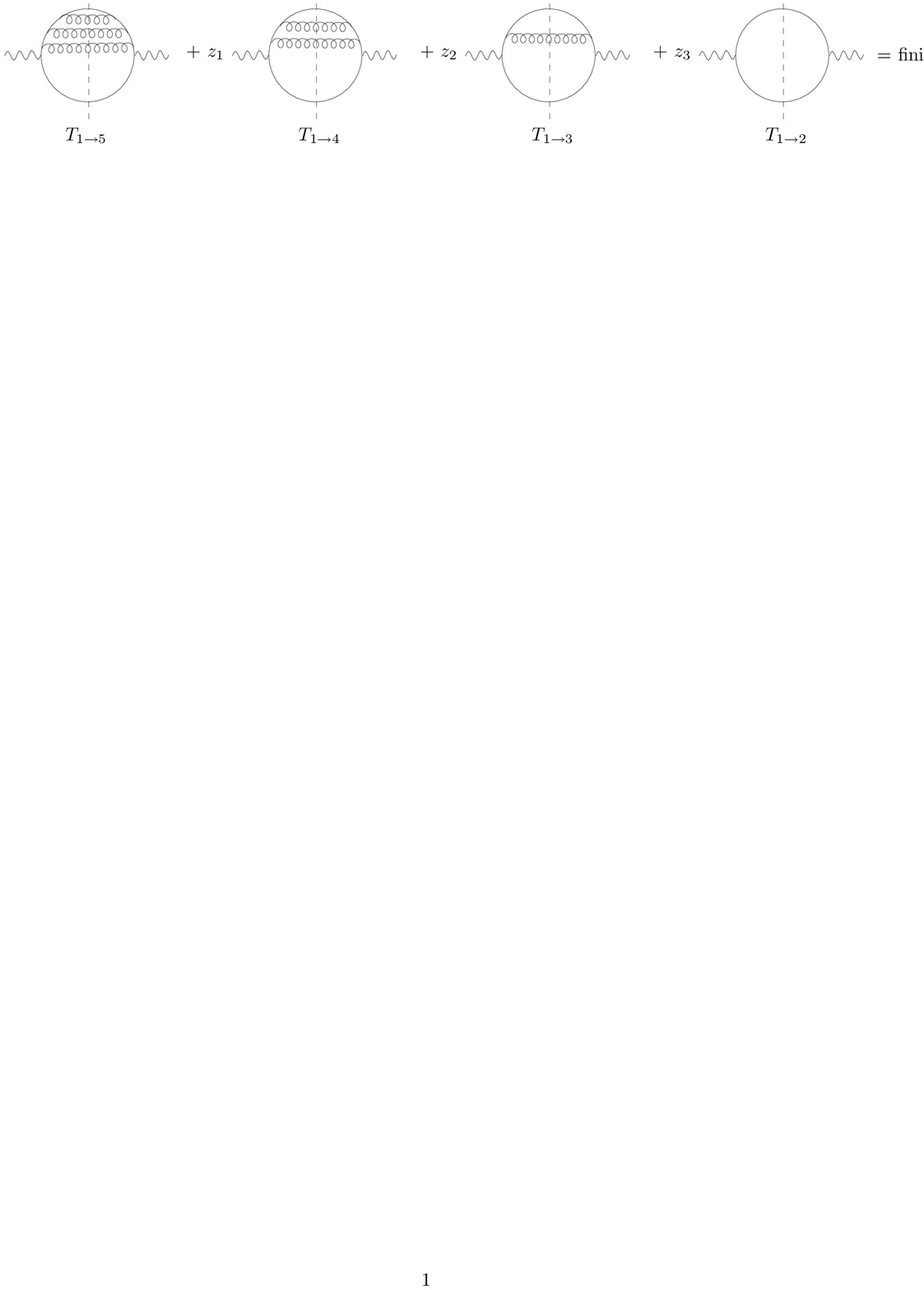,height=21.5cm}
%\end{center}
\vspace*{-19cm}
\caption{Cancellation of IR divergences in the sum over all cuts 
of the renormalised graph\label{fig2}}
\end{figure}

Summing over all cuts of this diagram and performing UV renormalisation, 
we obtain the condition 
\begin{eqnarray}
T_{1\to5}+z_1\,T_{1\to 4}+z_2\,T_{1\to 3}+z_3\,T_{1\to 2}&=&\mbox{finite}\;,
\label{cfin}
\end{eqnarray}
where $T_{1\to i}$ denotes the diagram with $i$ cut lines as shown in 
Figure~\ref{fig2}.

\subsection{UV renormalisation}
\label{sec:uv}
\begin{figure}[htb]
\begin{center}
\epsfig{file=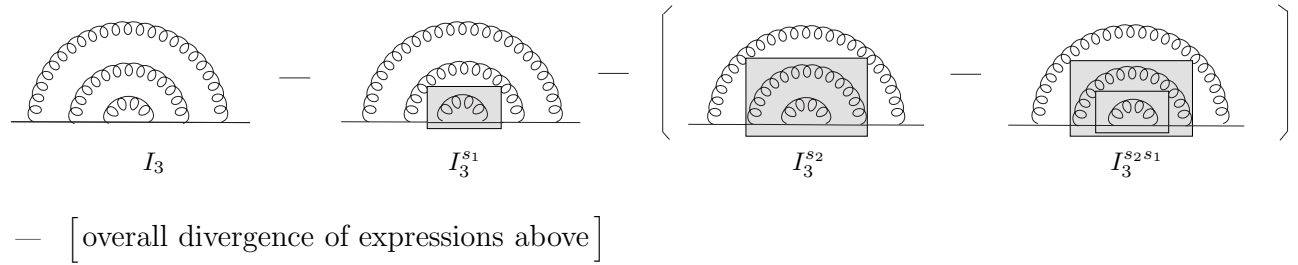,height=3.5cm}
\end{center}
\vspace*{-0.5cm}
\caption{UV renormalisation of the quark propagator at ${\cal
O}(\alpha_s^3)$\label{fig3}}
\end{figure}

For $i=1,2$, the renormalisation constants $z_i$ 
(in Feynman gauge) already have been 
calculated in \cite{Binoth:2004jv}, to be given by 
\begin{eqnarray}
z_1&=&C_F\frac{\as}{4\pi}\,\frac{1}{\eps}\\
z_2&=&C_F^2\left(\frac{\as}{4\pi}\right)^2\,
\left(\frac{1}{2\eps^2}-\frac{1}{4\eps}\right)
\end{eqnarray}
The 3-loop renormalisation constant $z_3$ will be derived in the
following.
Using the graphical BHPZ notation as in ref.\cite{Binoth:2004jv}, 3-loop
renormalisation of the fermion selfenergy implies 
that the combination of 
graphs as shown in Fig.~\ref{fig3} is finite. 

The explicit calculation yields:
\begin{eqnarray}
I_3&=&i\not p\,C_F^3\left(\frac{\as}{4\pi}\right)^3\,
\left(\frac{-p^2}{\mu^2}\right)^{-3\eps}
\frac{4(1-\eps)^3}{\Gamma^3(1+\eps)}\,G(1,1,0)G(1+2\eps,1,1)\nn\\
&\times&
\Bigl\{G(\eps,1,0)+G(1+\eps,1,0)-G(1+\eps,1,1) \Bigr\}\\
G(\alpha,\beta,n)&=&\frac{\Gamma(\alpha+\beta-D/2)}
{\Gamma(\alpha)\Gamma(\beta)}\,
%\mathrm{Beta}(D/2-\alpha+n,D/2-\beta)
\frac{\Gamma(D/2-\alpha+n)\,\Gamma(D/2-\beta)}{\Gamma(D-\alpha-\beta+n)}
\quad,\quad D=4-2\eps\nn\\
I_3^{s_1}&=&i\not p\,C_F^3\left(\frac{\as}{4\pi}\right)^3\,
\left(\frac{-p^2}{\mu^2}\right)^{-2\eps}
\frac{(1-\eps)^2\Gamma^3(1-\eps)\Gamma(1+2\eps)}
{\eps^3\Gamma^2(1+\eps)\Gamma(3-3\eps)}\\
I_3^{s_2}&=&i\not p\,C_F^3\left(\frac{\as}{4\pi}\right)^3\,
\left(\frac{-p^2}{\mu^2}\right)^{-\eps}
\frac{(1-\eps)\Gamma^2(1-\eps)}{\eps^3\Gamma(2-2\eps)}
\left\{\frac{1}{2}+\frac{5}{4}\eps-
\eps\log{(-p^2/\mu^2)} \right\}\\
I_3^{s_2s_1}&=&i\not p\,C_F^3\left(\frac{\as}{4\pi}\right)^3\,
\left(\frac{-p^2}{\mu^2}\right)^{-\eps}
\frac{(1-\eps)\Gamma^2(1-\eps)}{\eps^3\Gamma(2-2\eps)}
\left\{1+\eps-
\eps\log{(-p^2/\mu^2)} \right\}
\end{eqnarray}
The overall divergences $J^{s_i}_3=$\,overall div\,$[I^{s_i}_3]$ 
of the diagrams above are thus given by
\begin{eqnarray}
J_3&=&i\not p\,C_F^3\left(\frac{\as}{4\pi}\right)^3\,
\Bigl\{\frac{1}{6\eps^3}+\frac{3}{4\eps^2}-
\frac{1}{2\eps^2}\log{(-p^2/\mu^2)}\nn\\
&&+ 
\frac{1}{\eps}\,[\frac{79}{24}-\frac{\pi^2}{2}-
\frac{9}{4}\log{(-p^2/\mu^2)}+
\frac{3}{4}\log^2{(-p^2/\mu^2)}]  \Bigr\}
\label{ci3}\\
J_3^{s_1}&=&i\not p\,C_F^3\left(\frac{\as}{4\pi}\right)^3\,
\Bigl\{\frac{1}{2\eps^3}+\frac{5}{4\eps^2}-
\frac{1}{\eps^2}\log{(-p^2/\mu^2)}\nn\\
&&+   
\frac{1}{\eps}\,[\frac{31}{8}-\frac{\pi^2}{6}-
\frac{5}{2}\log{(-p^2/\mu^2)}+
\log^2{(-p^2/\mu^2)}]  \Bigr\}\\
J_3^{s_2}&=&i\not p\,C_F^3\left(\frac{\as}{4\pi}\right)^3\,
\Bigl\{\frac{1}{2\eps^3}+\frac{7}{4\eps^2}-
\frac{3}{2\eps^2}\log{(-p^2/\mu^2)}\nn\\
&&+  
\frac{1}{\eps}\,[\frac{9}{4}-\frac{\pi^2}{12}-
\frac{3}{2}\log{(-p^2/\mu^2)}+
\frac{5}{4}\log^2{(-p^2/\mu^2)}]  \Bigr\}\\
J_3^{s_2s_1}&=&i\not p\,C_F^3\left(\frac{\as}{4\pi}\right)^3\,
\Bigl\{\frac{1}{\eps^3}+\frac{2}{\eps^2}-
\frac{2}{\eps^2}\log{(-p^2/\mu^2)}\nn\\
&&+   
\frac{1}{\eps}\,[3-\frac{\pi^2}{6}-
3\log{(-p^2/\mu^2)}+
\frac{3}{2}\log^2{(-p^2/\mu^2)}]  \Bigr\}\;.
\label{ci321}
%\mathrm{pp}&=&\frac{-p^2}{\mu^2}\nn
\end{eqnarray}
Note that we have adopted the $\overline{\rm{MS}}$ prescription 
$$\alpha=C_{\overline{\rm{MS}}}\,\alpha^0, 
\as=C_{\overline{\rm{MS}}}\,\alpha_{s}^0\,,\;
C_{\overline{\rm{MS}}}=\Gamma(1+\eps)\,\left(\frac{4\pi}{\mu^2}\right)^\eps\;.$$
From eqs.~(\ref{ci3}) to (\ref{ci321}) we can now derive $z_3$ as
\begin{eqnarray}
i\not p\,z_3&=&
J_3-J_3^{s_1}-(J_3^{s_2}-J_3^{s_2s_1})\nn\\
\Rightarrow z_3&=&C_F^3\left(\frac{\as}{4\pi}\right)^3\,
\left(\frac{1}{6\eps^3}-\frac{1}{4\eps^2}+\frac{1}{6\eps}\right)\;.
\end{eqnarray}
The non-local logarithmic terms cancel, as guaranteed by the 
BHPZ theorem\cite{Bogoliubov:1957gp,Hepp:1966eg,Zimmermann:1969jj}.

\subsection{Combining the renormalised diagrams}
In order to verify eq.~(\ref{cfin}), we have to integrate the 
ladder diagrams 
corresponding to the process $\gamma^*\to i$\,partons  
over the $1\to i$ particle phase space. 
Up to $i=4$, this has been done already in ref.\cite{Binoth:2004jv}, where 
$T_{1\to 4}$ has been calculated by sector decomposition. The important 
new ingredient here is the calculation of $T_{1\to 5}$. Before showing this 
calculation in more detail, let us first construct the 
expressions entering eq.~(\ref{cfin}) for $i<5$. 
From ref.\cite{Binoth:2004jv}, we have
\begin{eqnarray}
T_{1\to 2} &=& 
%\int d\Phi_{1\to2} | {\cal M}_{1\to 2} |^2 
           2 \,\alpha\, q^2\, \left( \frac{q^2}{\mu^2}
	   \right)^{-\epsilon} \,
\frac{(1-\eps)\Gamma(1-\eps)}{\Gamma(1+\eps)\Gamma(2-2\eps)} \\
T_{1\to 3}&=&-z_1\, T_{1\to 2}\,\left( 
\frac{q^2}{\mu^2} \right)^{-\eps} 
 \frac{2\,(1-\eps)^2\Gamma(1-\eps)^2}
 {\Gamma(1+\eps)\Gamma(3-3\eps)}\\
T_{1\to 4}&=&(\eps \,z_1)^2\, T_{1\to 2}\,
\left( \frac{q^2}{\mu^2} \right)^{-2\eps}
 \frac{1}{\Gamma(1+\eps)^2\Gamma(1-2\eps)} \, 
 \left\{\frac{1}{2\eps^2} +\frac{11}{4\eps}+7.869\right\}
\end{eqnarray}
Combination of these results with the renormalisation constants 
given in section \ref{sec:uv} leads to
\begin{eqnarray}
&&z_1\,T_{1\to 4}+z_2\,T_{1\to 3}+z_3\,T_{1\to 2}=\label{c234}\\
&&C_F^3\left(\frac{\as}{4\pi}\right)^3\,T_{1\to 2}
\left\{\frac{1}{6\eps^3}+\frac{1}{2\eps^2}\,[3-\log{\left(\frac{q^2}{\mu^2}\right)}]
 +\frac{1}{\eps}\,[5.608-\frac{9}{2}\log{\left(\frac{q^2}{\mu^2}\right)]+
\frac{3}{4}\log^2{\left(\frac{q^2}{\mu^2}\right)}}\,+
\,\mbox{finite}\right\}\nn
\end{eqnarray}
What remains to be shown now is that the 5-parton contribution 
$T_{1\to 5}$ exactly cancels the poles in (\ref{c234}).

\subsection{Calculation of the 5-particle contribution}
The graph $T_{1\to 5}$ is calculated numerically by sector
decomposition. To this aim, the phase space integrals 
are brought to a form where all integrations are from zero to one, 
as described in more detail in appendix A. 
Note that the parametrisation given here is particularly convenient 
for the denominator structure of our sample topology. In order to deal 
with the full matrix element, several parametrisations have been worked out, 
each one optimised to be applied to a certain class of denominators. 
An automated subroutine scans the denominators of a given matrix element 
and applies the appropriate parametrisation. In this way, the full expression 
naturally is split into tractable subparts.

After having performed the transformations of the phase space integration 
variables  as explained in appendix \ref{app:a1}, the $1\to 5$ phase space in $D$
dimensions is given by eq.(\ref{psd}):
\begin{eqnarray}
\int d\Phi_{1\to 5}^D
&=&{\cal K}^{(5)}_\Gamma(q^2)^{2D-5}\int_0^1\prod\limits_{i=2}^{10} dt_i\,
[t_5(1-t_5)]^{-1-\eps}
[t_8(1-t_8)t_{10}(1-t_{10})]^{-\frac{1}{2}-\eps}\nn\\
&&
[t_2\,t_6(1-t_6)(1-t_7)]^{1-2\eps}
[(1-t_2)t_3(1-t_3)t_4(1-t_4)t_9(1-t_9)]^{-\eps}\,
t_7^{2-3\eps}\label{psdmain}\\
&&\nn\\
%{\cal K}^{(5)}_\Gamma=K^{(5)}_\Gamma*2^{-8\eps}
{\cal K}^{(5)}_\Gamma
&=&\frac{2\pi^{4\eps}}{(4\pi)^9\Gamma(-2\eps)\Gamma(2-2\eps)}\;.\nn
\end{eqnarray}
Singularities only occur at the boundaries $t_i=0,1$. Further, one can 
split the 
integrations at $t_i=1/2$ and remap the variables to the unit cube
to assure that all potential singularities  occur only for $t_i\to 0$. 
However, as this procedure doubles the number of integrals for each $t_i$, it  
is only done for those variables where a singularity at $t_i=1$ is possible at all,
in order to avoid a proliferation of terms. 

The matrix element typically contains terms of the structure  
%$$\frac{1}{t_6^2[t_6+t_9]}$$
\begin{eqnarray*}
I&=&
\int_0^1 dx\int_0^1 dy \,x^{-1-\epsilon}\,(x+y)^{-1}\;,
\end{eqnarray*}
where a naive subtraction of the singularity for $x\to 0$ of the form 
\begin{eqnarray*}
\int_0^1 dx\int_0^1 dy\, x^{-1-\eps} f(x,y)&=&-\frac{1}{\eps}\int_0^1 dy\,\
f(0,y)+\int_0^1 dx\int_0^1 dy\,x^{-\eps}\,\frac{f(x,y)-f(0,y)}{x}
\end{eqnarray*}
fails, because the singularities for $x\to 0$ and $y\to 0$ are overlapping. 
This is where sector decomposition shows its virtues. 
The working mechanism of sector decomposition already has been explained 
in detail in\cite{Binoth:2000ps} and therefore will be outlined only shortly here.
The basic idea is to first split the integration region into sectors 
where the variables $x$ and $y$ are ordered. 
$$I=\int_0^1 dx \int_0^1dy \,x^{-1-\epsilon}\,(x+y)^{-1}
\,[\underbrace{\Theta(x-y)}_{(1)}+\underbrace{\Theta(y-x)}_{(2)}]$$
Then remapping the 
integration domain  to the unit cube, the singularities 
in our simple example are already disentangled:
After the substitutions $y=x\,t$ in sector (1) and 
$x=y\,t$ in sector (2), one has
\begin{eqnarray*}
I&=&\int_0^1 dx\,x^{-1-\epsilon}\int_0^1 dt
\,(1+t)^{-1}+\int_0^1 dy
\,y^{-1-\epsilon}\int_0^1 dt\,t^{-1-\epsilon}\,(1+t)^{-1}\;.
\end{eqnarray*}
For more complicated functions, several iterations 
of this procedure may be necessary, but it is easily implemented 
into an automated subroutine. Once all singularities are factored out, 
the result can be 
expanded in $\epsilon$, where the subtraction of the pole terms 
naturally leads to plus distributions by the identity 
$$x^{-1+\kappa\epsilon}=\frac{1}{\kappa\,
\epsilon}\,\delta(x)+
\sum_{n=0}^{\infty}\frac{(\kappa\epsilon)^n}{n!}
\,\left[\frac{\ln^n(x)}{x}\right]_+\; \mbox{ where } 
\int_0^1 dx \, \left[f(x)/x\right]_+=\int_0^1 dx \, 
\frac{f(x)-f(0)}{x}\;.$$
In this way, a 
Laurent series in $\eps$ is obtained, where the pole coefficients are 
sums of finite parameter integrals which can be evaluated numerically. 

Note that the numerator structure of the matrix element can only 
improve the infrared pole structure, such that it can be included later,  
at the stage of the expansion in $\epsilon$. 
It also should be mentioned that for some phase space parametrisations, required 
to tackle the full matrix element, square-root terms in the denominator are 
unavoidable. Such terms can spoil the simple scaling behaviour which is 
crucial for the algorithm to work. However, one can always find  (nonlinear) 
variable transformations such that these terms can be mapped to a form
which is amenable to sector decomposition.

Applying the method to our sample diagram, and 
requiring a numerical precision of 1\%, the following result 
is obtained after an integration time of about 20 minutes on 
a 2.8 GHz Pentium IV PC:
\begin{eqnarray}
T_{1\to 5}
%&=&C_F^3\left(\frac{\as}{4\pi}\right)^3\,T_{1\to 2}
%\left(\frac{Q^2}{\mu^2}\right)^{-3\eps}
%\frac{(1-\eps)^3 4^{-4\eps}}{16\pi^2\Gamma(1+\eps)^3
%\Gamma(1-\eps)\Gamma(-2\eps)}\left\{
%\frac{13.1556}{\eps^4}+\frac{230.795}{\eps^3}+
%\frac{2040.0}{\eps^2}+\frac{12147.1}{\eps}
%\right\}\nn\\
&=&-C_F^3\left(\frac{\as}{4\pi}\right)^3\,T_{1\to 2}
\left\{
\frac{0.16662}{\eps^3}+\frac{1}{\eps^2}\,
[1.4993-0.4999\,\log{\left(\frac{q^2}{\mu^2}\right)}]\right.\nn\\
&&\left.+\frac{1}{\eps}\,[ 5.5959-4.4978\,\log{\left(\frac{q^2}{\mu^2}\right)}+
0.74978\,\log^2{\left(\frac{q^2}{\mu^2}\right)} ]\,+\,\mbox{finite}
\right\}\;.
\label{t5}
\end{eqnarray}
Combining eqs. (\ref{c234}) and (\ref{t5}) we see that all poles
cancel within the numerical precision. 

\section{Inclusion of a measurement function}
\setcounter{equation}{0}
\label{sec:num}

The isolation of infrared poles by sector decomposition is an algebraic 
procedure, leading to a set of finite functions for each pole coefficient 
as well as for the finite part. The finite part can be written to a 
Monte Carlo program and combined with any infrared safe measurement function. 
To this aim, one has to take the limit $D\to 4$ of the $D$-dimensional 
phase space, which is non-trivial because in $D=4$, the Gram determinant 
of five light-like momenta vanishes, which means that only 8 Mandelstam 
invariants are independent, 
whereas in $D=4-2\eps$ one has 9 independent invariants, i.e. 
9 independent phase space integration variables, 
and sector decomposition acts in $D=4-2\eps$ dimensions. 
How this problem is solved is explained in appendix \ref{app:a2}. 
It is also described there how  
the four-momenta of the particles in the final state in terms of 
energies and angles  
are reconstructed from the phase space integration variables $t_i$. 
Note that the variables $t_i$ are transformed in the course of sector
decomposition, such that for each function which is an endpoint of the 
sector decomposition tree, the expressions for the invariants
$s_{ij}$ in terms of the final Monte Carlo integration variables look 
different. This requires careful (automated) book-keeping, but does not 
constitute a principal problem. 

Further, it has to be assured that the subtraction terms only come to 
action in phase space regions which are allowed by the measurement function. 
To illustrate this point, consider the simple one-dimensional example 
where the measurement 
function is just a  step function $\Theta(x-a),\, a>0$, and the 
``matrix element" 
after sector decomposition is given by a plus distribution $[f(x)/x]_+$. 
If we naively combine the plus distribution with our measurement 
function, we obtain
\begin{equation}
\int_0^1\, dx \,\frac{f(x)-f(0)}{x}\,\Theta(x-a)=f(0)\,\ln{a}+
\int_a^1\, dx \,\frac{f(x)}{x}\;.
\label{nosubt}
\end{equation}
On the other hand, the 
$f(0)$ term stems from the subtraction of a singularity
at $x=0$, which is now killed by our measurement function anyway, 
such that inclusion of the $f(0)$ term would lead to a wrong result.
Therefore, the correct way to include the measurement function is 
of course 
\begin{equation}
\int_0^1 dx \,\frac{f(x)\Theta(x-a)-f(0)\Theta(-a)}{x}\;.
\end{equation}
However, this does {\em not} mean that the $\eps$--expansions and 
subtractions have to be redone each time the measurement function is changed. 
It can be achieved by including symbolic functions in the 
$\eps$--expansion
which, depending on the generated phase space point in the 
Monte Carlo program, 
take on the appropriate values. 

\begin{figure}[htb]
\begin{center}
\epsfig{file=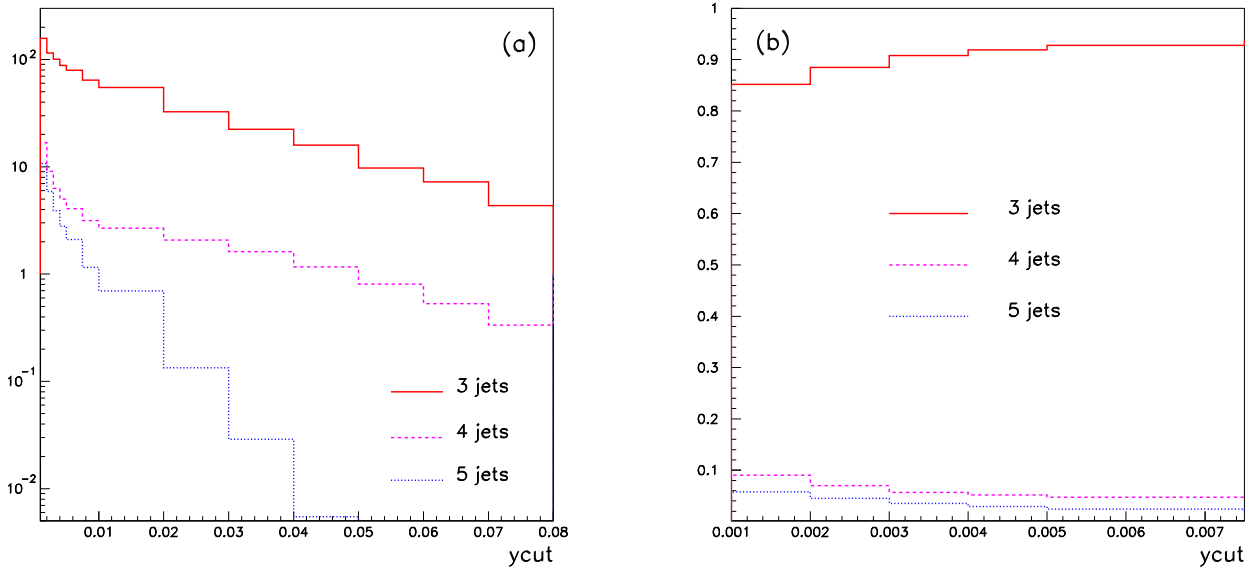,height=7.5cm}
\end{center}
\vspace*{-5mm}
\caption{3--, 4-- and 5--jet rates at order $\alpha_s^3$ for  
the sample matrix element
\label{fig4}}
\end{figure}

As an example, the JADE algorithm\cite{Bethke:1988zc} to 
define 3--, 4-- and 5--jet events 
has been implemented into a Monte Carlo program built upon 
the output of sector decomposition, using the 
multi-dimensional integration package BASES\cite{Kawabata:1995th}.  
For the plots in Figs.~\ref{fig4}a and \ref{fig4}b, 
the diagram  discussed in this paper (summed over all cuts) 
served as a toy matrix element, but it should be emphasised  
that the same Monte Carlo program can be used to calculate the 
full process $e^+e^-\to 3,4,5$\,jets at order $\alpha_s^3$, 
once the contributions from 
the other topologies are implemented. 
Further, the architecture of the program is such that the JADE 
algorithm can be easily replaced by a different jet algorithm, and shape 
observables can also be defined. 

%\begin{wrapfigure}[10]{r}[0.cm]{5.8cm}
%\vspace*{-1.2cm}
%\epsfxsize=5.8cm   %width of figure - will enlarge/reduce the figures
%\epsfbox{rain345jets.eps}
%\caption{3--, 4-- and 5--jet rates at order $\alpha_s^3$\label{fig4a}}
%\end{wrapfigure}

Figures \ref{fig4}a and \ref{fig4}b show the 3--, 4-- and 5--jet rates 
at order $\alpha_s^3$, as a function of the 
jet separation parameter $y^{\rm cut}$. In Figure \ref{fig4}a, the y-axis 
is in arbitrary units, whereas in Figure \ref{fig4}b the rates are 
normalised to the sum of the 3--, 4-- and 5--jet rates. The renormalisation
scale $\mu^2$ has been set equal to $q^2$, the center of mass energy of the 
$e^+e^-$ system. As in the previous section, the numerical precision is 1\%. 
Figure \ref{fig4}a shows that the 5--jet rate drops drastically as $y^{\rm cut}$ 
increases, as to be expected. The 3-jet rate decreases only slowly, as only few 
events are classified as 2--jet events and thus rejected for values of 
 $y^{\rm cut}\leq 0.08.$ 
Figure \ref{fig4}b demonstrates how the 3-jet rate decreases in favour of the 
4-- and 5--jet rates if $y^{\rm cut}$ becomes very small.

%\clearpage

\section{Conclusions and Outlook}
\label{sec:conc}
In this paper, a method based on sector decomposition to calculate 
the double real radiation part of the process 
$e^+e^-\to 3$\, jets at ${\cal O}(\as^3)$ has been presented. 
The sector decomposition  algorithm serves to 
%algebraically automatically
isolate, by an automated algebraic subroutine, the 
infrared poles which occur upon phase space integration if one or 
several particles in the final state become soft and/or collinear.
In this way, one is dispensed from 
the manual construction of a subtraction scheme. The cancellation of the 
poles is shown by numerical calculation of the pole coefficients. 

For the process $e^+e^-\to 3$\, jets at NNLO, integration over a 
phase space with up to five particles in the final state is necessary, 
where up to two particles can become 
soft and collinear. It has been proven that the program handles the isolation 
and subtraction of the poles  correctly by considering all possible cuts of a 
specific diagram which is a 
subpart of the colour structure $\sim C_F^3$ contained in the 
full matrix element at order $\as^3$. Summing over all the cuts, the poles
cancel within the numerical precision.   

The finite part has been implemented into a Monte Carlo program which 
allows the inclusion of a measurement function in order to obtain 
differential distributions for arbitrary (infrared safe) observables. 
As an example, the  3--, 4-- and 5--jet rates 
at order $\alpha_s^3$ as a function of the 
jet separation parameter $y^{\rm cut}$ are shown for the subpart of the 
full matrix element treated in this paper. 
As this toy matrix element 
already shows most of the problems which occur in the double real 
radiation part, while the one-loop virtual corrections 
combined with the $1\to 4$ phase space, as 
well as the 
two-loop virtual part combined with the $1\to 3$ phase space,  
are relatively easy (because the virtual integrals only lead to 
renormalisation factors), it is an ideal testing ground 
for the method presented here to tackle massless $1\to 5$ processes. 
For the calculation of the full matrix element, 
the expressions to be integrated numerically in the double real radiation part 
will of course be much larger, but the method described in this paper 
can handle them in the same way.  
The  virtual corrections  will also be 
more complicated, but can  be treated with sector decomposition 
applied to loop integrals\cite{Binoth:2000ps}, 
for the kinematics of $e^+e^-$ annihilation. 
Therefore the problem is basically reduced to large file handling, 
book-keeping and implementation time/CPU time.

For the parts of the full matrix element considered so far, the numerical 
stability is very good. A reason might be that the subtractions within the 
sector decomposition method are local in the sense of plus distributions,  
i.e. the singular limits in each integration variable 
are directly subtracted.

CPU time will become an issue for the treatment of the full process, 
but as the method relies on a division of the amplitude squared 
into different ``topologies" corresponding to different classes of 
denominator structures, the problem is naturally split into smaller 
subparts. If such a ``trivial parallelisation" is not sufficient, 
there is still the possibility to parallelise the evaluation 
of the functions produced by sector decomposition. 

As the method is based on a universal algorithm acting on integration 
variables, and does not require analytic integration over 
complicated functions, it will surely see a number of interesting applications
in the future, in particular in what concerns the production of 
massive particles.

\section*{Acknowledgements}
I would like to thank Thomas Gehrmann for fruitful discussions on the subject
and for reading the manuscript. 
This work was supported in part by the Swiss National Science Foundation
(SNF) under contract number 200020-109162.

\begin{appendix}
\renewcommand{\theequation}{\mbox{\Alph{section}.\arabic{equation}}}
\section{Massless 5-parton Phase Space}
\setcounter{equation}{0}
\label{app:ps}

\subsection{Phase space $1\to 5$ in $D\not=4$ dimensions}
\label{app:a1}

The phase space for the decay of one off-shell particle 
with momentum $q$ into $N$ massless particles with momenta 
$p_1,\ldots,p_N$ in $D$ dimensions is given by
\begin{eqnarray}
\int d\Phi_{1\to N}^D&=&
(2\pi)^{ N - D (N-1)} \int  \prod\limits_{j=1}^{N} d^Dp_j \,\delta^+(p_j^2) 
\delta\Bigl(q-\sum\limits_{i=1}^{N} p_i \Bigr)\nonumber\\
&=&(2\pi)^{ N - D (N-1)}\, 2^{1-N}\int \prod\limits_{j=1}^{N-1} 
d^{D-1} \vec{p}_j\,\frac{\Theta(E_j)}{E_j}
 \, \delta^+([q-\sum\limits_{i=1}^{N-1} p_i]^2)\;.
 \label{psgen}
\end{eqnarray}
For $N=5$, we 
parametrise the momenta in $D$ dimensions as 
(ordering of the vector components:\\ $(E,(D-4),x,y,z)$)
\begin{eqnarray}
q&=&(q,\vec{0}^{(D-1)})\nn\\
p_1&=&E_1\,(1,\vec{0}^{(D-2)},1)\nn\\
p_2&=&E_2\,(1,\vec{0}^{(D-3)},\sin\theta_1,\cos\theta_1)\nn\\
p_3&=&E_3\,(1,\vec{0}^{(D-4)},\sin\theta_2\sin\theta_4,\sin\theta_2\cos\theta_4,\cos\theta_2)\nn\\
p_4&=&E_4\,(1,(\vec{0}^{(D-5)},\sin\theta_6 \sin\theta_5\sin\theta_3),
\cos\theta_6\sin\theta_5\sin\theta_3,\cos\theta_5\sin\theta_3,\cos\theta_3)\nn\\
p_5&=&q-p_1-p_2-p_3-p_4\label{pddim}
\end{eqnarray}
Inserting this parametrisation into (\ref{psgen}) 
and carrying out integrations over azimuthal angles 
leads to
\begin{eqnarray}
\int d\Phi_{1\to 5}^D&=&
(2\pi)^{5-4D} 2^{-4} V(D-1)V(D-2)V(D-3)V(D-4)\nn\\
&&\int  \prod\limits_{j=1}^{4} dE_j \,\Theta(E_j)\,d\theta_1\ldots d\theta_6
\left[E_1E_2E_3E_4\sin\theta_1\sin\theta_2\sin\theta_3\right]^{D-3} \nonumber\\
&&
\left[\sin\theta_4\sin\theta_5\right]^{D-4}(\sin\theta_6)^{D-5}
\, \delta^+([q-\sum\limits_{i=1}^{4} p_i]^2)\label{ps5}\\
&&\nonumber\\
\mbox{where }\,V(D)&=&2\pi^{\frac{D}{2}}/\Gamma(\frac{D}{2})\;.\nonumber
\end{eqnarray}
Introducing the scaled invariants $y_i$ as new integration variables 
\begin{eqnarray*}
&&y_1=s_{12}/q^2, y_2=s_{13}/q^2, 
y_3=s_{23}/q^2,
y_4=s_{14}/q^2,y_5=s_{24}/q^2,\\
&&y_6=s_{34}/q^2, 
y_7=s_{15}/q^2, y_8=s_{25}/q^2, 
y_9=s_{35}/q^2,
y_{10}=s_{45}/q^2
\end{eqnarray*}
leads to the Jacobian
\begin{eqnarray*}
|\det J|&=&2^{10}q^4\left[E_1E_2E_3E_4\sin\theta_1\sin\theta_2\sin\theta_3\right]^{3}
\left[\sin\theta_4\sin\theta_5\right]^{2}\sin\theta_6\;.
\end{eqnarray*}
The Jacobian can be expressed in terms of the 
determinant of the Gram matrix $G_{ij}=2p_i\cdot p_j$
\begin{eqnarray*}
\det G&=&-2^{5}q^2\left[E_1E_2E_3E_4\sin\theta_1
\sin\theta_2\sin\theta_3\sin\theta_4\sin\theta_5\sin\theta_6\right]^{2}\\
\Rightarrow |\det J|&=&\sqrt{2^{15}q^6\,(-\det G)}\,\left[E_1E_2E_3E_4\sin\theta_1
\sin\theta_2\sin\theta_3\right]^{2}\sin\theta_4\sin\theta_5\;.
\end{eqnarray*}
After these variable transformations, the phase space is given by
\begin{eqnarray}
\int d\Phi_{1\to 5}^D&=&
(2\pi)^{5-4D}2^{-2-2D}V(D-1)V(D-2)V(D-3)V(D-4)(q^2)^{2D-5}\nn\\
&&\int\prod\limits_{j=1}^{10} dy_j\,
\delta(1-\sum\limits_{i=1}^{10} y_i)\,
(-\Delta_5)^{\frac{D}{2}-3}\Theta(-\Delta_5)\label{psddim}
\end{eqnarray}
where
\begin{eqnarray}
-\Delta_5&=&y_{10}^2y_1y_2y_3+y_9^2y_1y_4y_5+y_8^2y_2y_4y_6+y_7^2y_3y_5y_6+y_6^2y_1y_7y_8\nn\\
&&+y_5^2y_2y_7y_9+y_4^2y_3y_8y_9+y_3^2y_4y_7y_{10}+y_2^2y_5y_8y_{10}+y_1^2y_6y_9y_{10}\nn\\
&&+y_{10}\,[y_2y_3y_5y_7+y_1y_3y_6y_7+ y_2y_3y_4y_8+ y_1y_2y_6y_8+ y_1y_3y_4y_9+
y_1y_2y_5y_9 ]\nn\\
&&+y_9\,[y_4y_5(y_3y_7+y_2y_8)+y_1y_6(y_5y_7+y_4y_8)]+y_6y_7y_8(y_3y_4+y_2y_5)\nn\\
&=&-\frac{1}{2}\,\det G/(q^2)^5\;.
\label{del5}
\end{eqnarray}
Note that $V(D-4)=2\pi^{-\eps}/\Gamma(-\eps)={\cal O}(\eps)$ is  
compensated by a spurious pole from $(-\Delta_5)^{-1-\eps}$.

For the phase space integration over the full 
$1\to 5$ matrix element relevant for 
the calculation of $e^+e^-\to 3$\,jets at NNLO, we 
choose different parametrisations optimised for 
certain types of denominators occurring in the matrix element.
In the following we only give the parametrisation which is relevant for 
the topology under consideration in this article.
In this parametrisation, we eliminate 
$y_1$ by $\delta(1-\sum\limits_{i=1}^{10} y_i)$
and substitute $y_6$ and $y_7$ in favour of 
$x_6=s_{134}/q^2$, $t_7=s_{1345}/q^2$. After these substitutions, 
the constraint 
$\Theta(-\Delta_5)$ is solved for $y_5$, leading to 
$$y_5^{\pm}=y_5^0\pm \sqrt{R_{5}}\;.$$ 
Then the condition $R_{5}\geq 0$ is solved for $y_8$ and the 
condition $(y_8^+-y_8^-)\geq 0$ is solved for $y_{10}$. 
Making variable transformations such that all integration limits 
over the new variables $t_i$ are from zero to one, we finally obtain
\begin{eqnarray}
s_{1345}/q^2&=&t_7\nn\\
s_{134}/q^2&=&t_6\,t_7\nn\\
s_{13}/q^2&=&t_6\,t_7\,(1-t_{2})\nn\\
s_{23}/q^2&=&t_3\,(1-t_7)(1-t_2t_4)(t_6\,(1-t_9)+t_9)\nn\\
s_{14}/q^2&=&t_2\,t_4\,t_6\,t_7\nn\\
s_{24}/q^2&=&y_5^-+(y_5^+-y_5^-)\,t_5\nn\\
s_{34}/q^2&=&t_2\,t_6\,t_7\,(1-t_4)\nn\\
s_{15}/q^2&=&t_7\,(1-t_6)\,[1-t_9\,(1-t_2t_4)]-y_{10}\nn\\
s_{25}/q^2&=&y_8^-+(y_8^+-y_8^-)\,t_8\nn\\
s_{35}/q^2&=&t_7\,t_9\,(1-t_6)(1-t_2t_4)\nn\\
s_{45}/q^2&=&y_{10}^-+(y_{10}^+-y_{10}^-)\,t_{10}
\label{traf}\\
&&\nn\\
y_8^\pm&=&y_8^0\pm d_8/2\nn\\
y_8^0&=&(1 - t_6)\,(1 - t_7)\,\{t_9 + t_3\,[t_6\,(1 - t_9) - t_9]\}/
     (t_6\,(1 - t_9) + t_9)\nn\\
d_8&=&y_8^+ -y_8^-=4\,(1-t_6)\,(1-t_7)\,
\sqrt{(1 - t_3)\,t_3\,t_6\,(1 - t_9)\,t_9}/
        (t_6\,(1-t_9) + t_9)\nn\\
y_{10}^\pm&=&y_{10}^0\pm d_{10}/2\nn\\
y_{10}^0&=&t_2\,t_7\,(1 - t_6)\,
\{1 - t_9 - t_4\,[1 -t_9 (2 - t_2)\,]\}/
      (1 - t_2\,t_4)\nn\\
d_{10}&=&y_{10}^+-y_{10}^-=4\,t_7\,t_2\,(1-t_6)\,
\sqrt{(1 - t_2)\,(1 - t_4)\,t_4\,
         (1 - t_9)\,t_9}/(1 - t_2\,t_4)\;.
\nn
\end{eqnarray}  
The solution of $\Delta_5=0$, $y_5^\pm$, is rather lengthy and 
thus will not be given explicitly.

In terms of the new variables, the phase space is given by
\begin{eqnarray}
\int d\Phi_{1\to 5}^D
&=&{\cal K}^{(5)}_\Gamma(q^2)^{2D-5}\int_0^1\prod\limits_{j=2}^{10} dt_j\,
[t_5(1-t_5)]^{-1-\eps}
[t_8(1-t_8)t_{10}(1-t_{10})]^{-\frac{1}{2}-\eps}\nn\\
&&
[t_2\,t_6(1-t_6)(1-t_7)]^{1-2\eps}
[(1-t_2)t_3(1-t_3)t_4(1-t_4)t_9(1-t_9)]^{-\eps}\,
t_7^{2-3\eps}\label{psd}\\
&&\nn\\
%{\cal K}^{(5)}_\Gamma=K^{(5)}_\Gamma*2^{-8\eps}
{\cal K}^{(5)}_\Gamma&=&(2\pi)^{5-4D}2^{-2-2D}2^{-8\eps}
V(D-1)V(D-2)V(D-3)V(D-4)\nn\\
&=&\frac{\pi^{4\eps}}{2^{17}\pi^9\Gamma(-2\eps)\Gamma(2-2\eps)}\;.
\nn
\end{eqnarray}

\subsection{Phase space $1\to 5$ in $D=4$ dimensions}
\label{app:a2}

In $D=4$ dimensions, the Gram determinant $\Delta_5$ is 
zero due to the fact that already 4 independent 
light-like momenta $p_i\in \{p_1,\ldots,p_5\}$ span Minkowski space.
This leads to a nonlinear constraint between the Mandelstam 
variables $y_i$, as can be seen from eq.~(\ref{del5}).
The momenta in $D=4$ can parametrised as in eq.~(\ref{pddim}),  
but with $\theta_6=0$ (and no $(D-4)$-dimensional component). 
The constraint $\Theta(-\Delta_5)$ in eq.~(\ref{psddim}) becomes 
$\delta(-\Delta_5)$, leading to $y_5=y_5^\pm$ instead of 
$y_5^-+(y_5^+-y_5^-)\,t_5$, 
that is, $t_5$ takes only the values 0 or 1. 
Therefore, a consistent way to obtain the 4-dimensional phase space 
from the $D$-dimensional one is to integrate over $t_5$ 
 in (\ref{psd})  {\it before} any sector decomposition 
is performed, cancelling the spurious pole coming from the $t_5$ integration 
with $V(D-4)$ contained in ${\cal K}^{(5)}_\Gamma$. The matrix element ME 
will always be of the form ME$=A_0+A_1\,y_5+A_2\,y_5^2$ because in all cases 
where $y_5$ is in the denominator, a different parametrisation will be chosen, 
such that the same arguments hold for a different invariant $y_i$ 
with $i\not=5$.
Using the fact that in our case\footnote{The generalisation to the case $A_2\not=0$ is trivial, leading only 
to additional $\Gamma$ functions.} $A_2=0$
and writing $y_5$ as $y_5=y_5^+\,t_5+y_5^-\,(1-t_5)$ we obtain
\begin{eqnarray}
&&\int d\Phi_{1\to 5}^D\,{\rm{ME}}=\nn\\
&&R^{(5)}_\Gamma\,(q^2)^{2D-5}
\int_0^1\prod\limits_{j=2}^{10} dt_j\,
\left\{[A_0+y_5^+A_1]+
[A_0+y_5^-A_1]\right\}
[t_8(1-t_8)t_{10}(1-t_{10})]^{-\frac{1}{2}-\eps}\nn\\
&&
[t_2\,t_6(1-t_6)(1-t_7)]^{1-2\eps}
[(1-t_2)t_3(1-t_3)t_4(1-t_4)t_9(1-t_9)]^{-\eps}\,
t_7^{2-3\eps}\label{psd2}\\
&&\nn\\
%{\cal K}^{(5)}_\Gamma=K^{(5)}_\Gamma*2^{-8\eps}
&&R^{(5)}_\Gamma={\cal K}^{(5)}_\Gamma\,
\frac{\Gamma(-\eps)\,\Gamma(1-\eps)}{\Gamma(1-2\eps)}
%{\rm Beta}(-\eps,1-\eps)=
\frac{(2\pi)^{4\eps}}{(4\pi)^8\Gamma^2(1/2-\eps)\Gamma(2-2\eps)}
\end{eqnarray}
The new prefactor $R^{(5)}_\Gamma$ is finite in the limit $\eps\to 0$. 
The matrix element only depends on the eight independent variables 
$t_2,\ldots,t_4,t_6,\ldots,t_{10}$ now and sector decomposition in those variables 
will isolate the ``physical" infrared poles. 
Therefore we can still use the parametrisation (\ref{traf}) in $D=4$, the only 
difference being that $s_{24}/q^2$ is given by $y_5^+$ respectively $y_5^-$.

In order to construct a Monte Carlo program of (partonic) event generator type,
it is useful to express the four momenta again in terms of angles and energies. 
The corresponding expressions in terms of Mandelstam variables are
\begin{eqnarray}
E_1&=&\frac{q^2-s_{2345}}{2q},\;
E_2=\frac{q^2-s_{1345}}{2q},\;
E_3=\frac{q^2-s_{1245}}{2q},\;
E_4=\frac{q^2-s_{1235}}{2q}\nn\\
&&\nn\\
\cos\theta_1&=&-1+2\,(s_{1345}s_{2345}-s_{345})/(1-s_{1345})/(1-s_{2345})\nn\\
\cos\theta_2&=&-1+2\,(s_{1245}s_{2345}-s_{245})/(1-s_{1245})/(1-s_{2345})\nn\\
\cos\theta_3&=&-1+2\,(s_{1235}s_{2345}-s_{235})/(1-s_{1235})/(1-s_{2345})\nn
%\sin\theta_1&=&2\sqrt{s_{12}(s_{1345}s_{2345}-s_{345})}/(1-s_{1345})/(1-s_{2345})\nn\\
%\sin\theta_2&=&2\sqrt{s_{13}(s_{1245}s_{2345}-s_{245})}/(1-s_{1245})/(1-s_{2345})\nn\\
%\sin\theta_3&=&2\sqrt{s_{14}(s_{1235}s_{2345}-s_{235})}/(1-s_{1235})/(1-s_{2345})
\end{eqnarray}
The expression for $\cos\theta_4$ and $\cos\theta_5$ are more complicated and 
will not be given explicitly.

Taking the limit $\eps\to 0$ in (\ref{psd2}), the phase space integral over 
a matrix element ME in the above parametrisation is given by
\begin{eqnarray}
\int d\Phi_{1\to 5}^{D=4}\,{\rm{ME}}&=&\frac{(q^2)^3}{4^8\pi^9}
\int\prod\limits_{j=2, j\not= 5}^{10} dt_j\,
\left\{{\rm{ME}}\Big|_{t_5=0}+
{\rm{ME}}\Big|_{t_5=1}\right\}\nn\\
&&[t_8(1-t_8)t_{10}(1-t_{10})]^{-\frac{1}{2}}
\,t_2\,t_6\,t_7^2\,(1-t_6)(1-t_7)\;.
\label{ps4}
\end{eqnarray}

\end{appendix}

\end{document}